\documentclass[twocolumn,nofootinbib,showpacs,preprintnumbers,aps]{revtex4}
\usepackage {amsfonts}
\usepackage {graphicx}
\usepackage {longtable}
\begin{document}

\title{First observation of spin dichroism with deuterons up to 20 MeV in a carbon target}

\author{V.~Baryshevsky\protect\footnote{
E-mail: bar@inp.minsk.by; v\_baryshevsky@yahoo.com},
A.~Rouba\protect\footnote{ E-mail: rouba@inp.minsk.by}}
\address{Research Institute of Nuclear Problems, Bobruiskaya Str.\@11, 220050 Minsk, Belarus}
\author{R.~Engels, F.~Rathmann, H.~Seyfarth, H.~Str\"oher, T.~Ullrich}
\address{Institut f\"ur Kernphysik, Forschungszentrum J\"ulich,
  Leo-Brandt-Str.\@1, 52425 J\"ulich, Germany}
\author{C.~D\"uweke, R.~Emmerich, A.~Imig, J.~Ley, H.~Paetz~gen.~Schieck, R.~Schulze, G.~Tenckhoff, C.~Weske}
\address{Institut f\"{u}r Kernphysik, Universit\"at zu K\"oln,
Z\"ulpicher Str.\@77, D-50937 K\"oln, Germany}
\author{M.~Mikirtytchiants, A.~Vassiliev}
\address{PNPI, 188300 Gatchina, Russia}

%\begin{center} {\bf \today}
%\end{center}

%\begin{center}

\begin{abstract}

The first observation of the phenomenom of deuteron spin dichroism
in the energy region of 6-20~MeV is described. Experimental values
of this effect for deuterons after passage of an unpolarized
carbon target are reported.
\end{abstract}
%\end{center}

\maketitle
%\twocolumn

\section{INTRODUCTION}

It has been a longtime belief that the phenomena of rotation of
the polarization plane and birefringence applies to only photons.
The discovery of neutron spin precession in a polarized target due
to the nuclear pseudomagnetic field \cite{bar64}-\cite{abr82} and
subsequent analyses have shown that analogues of these effects
exist not only for photons, but also for other particles
\cite{bar92,bar93}. Thus, the optical effects caused by the
anisotropy of matter  represent only a special case of the
coherent phenomena arising at the passage of polarized particles
through matter.

The investigation of spin-dependent interactions of particles
is an important part of the research programs
to be carried out at storage rings, e.g. RHIC or COSY.
The study of a number of phenomena in particle physics
demands knowledge of imaginary and real parts of the forward
scattering amplitude.
Whereas the imaginary part, according to the optical theorem, can
be  found through the total scattering cross section, the real
part may be obtained through the dispersion relations or by using
the extrapolation of results from small-angle scattering.
Thus, the phenomena of birefringence (spin rotation and
oscillation) and spin dichroism (defined as the production of spin
polarization in an unpolarized beam) of deuterons
\cite{bar92,bar93} moving through homogeneous isotropic matter are
of great interest.
These phenomena allow the measurement of the real part
of the spin-dependent forward scattering amplitude directly.
Thus,
 a check of the dispersion relations on the basis of independent
measurements of the imaginary and real part becomes possible.
The confirmation of the existence of these effects would
necessitate taking them into account in all experiments where
particles with spins higher than 1/2 are scattered from polarized
and unpolarized targets.
Especially, their contribution to experiments in storage rings
\cite{bar02} and deuteron polarization measurements with thick
targets \cite{kox} has to be considered.
Particularly, the considered effect in the residual gas of a
storage ring should be taken into account for precision EDM
measurements \cite{EDM}.

The experiment analyzed in this paper has been devoted to the
first attempt to measure spin dichroism, i.\@ e.\@ creation of
tensor polarization in an unpolarized deuteron beam by unpolarized
carbon targets. The experiment was carried out in the Institute of
Nuclear Physics of the University of Cologne using the
electrostatic HVEC tandem Van-de-Graaff accelerator with deuterons
of up to 20~MeV. A $^3$He polarimeter, based upon the reaction $d\
+\ ^3He \rightarrow ^4He\ +\ p$, was used to measure polarization
of transmitted deuterons.
\bigskip

\section{THEORY: ROTATION AND OSCILLATION OF DEUTERON SPIN IN UNPOLARIZED
MATTER (BIREFRINGENCE AND SPIN DICHROISM)}

According to \cite{bar92,bar93} the index of refraction for a
deuteron (spin S=1) can be written as:

\begin{equation}
\label{hatn1}
\hat {N} = 1 + \frac{{2\pi \rho} }{{k^{2}}}\hat {f}\left( {0} \right),
\end{equation}
where$\hat {f}\left( {0} \right) = Tr\hat {\rho} _{J} \hat
{F}\left( {0} \right)$, $\rho $~is the density of matter (the
number of scatterers in 1 cm$^3$), $k$~is the deuteron wave
number, $\hat {\rho} _{J} $~is the spin density matrix of the
scatterers, $\hat {F}\left( {0} \right)$~is the operator of the
forward scattering amplitude acting in the combined spin space of
the deuteron and scatterer spin $\vec {J}$.

For an unpolarized target $ \hat {f}(0)$ can be written as:
\begin{equation}
\label{hatf} \hat {f}\left( {0} \right) = d + d_{1} S_{z}^{2}.
\end{equation}
where $S_z$~is the z component of the deuteron spin operator. The
axis of quantization \textit{z} is directed along
%$\vec {n} =\frac{{\vec {k}}}{{k}}$, where \textit{}
the particle wave vector $\vec {k}$.
Considering only strong interactions, which are invariant to the
parity transformation and time reversal, we may omit the terms
containing $S$ in odd degrees.
Therefore, the refractive index for deuterons
\begin{equation}
\label{hatn2}
 \hat {N} = 1 +
\frac{{2\pi \rho} }{{k^{2}}}\left( {d + d_{1} S_{z}^2}  \right)
\end{equation}
depends on the deuteron spin orientation relative to the deuteron
momentum.

The refractive index for a particle in the state, which is the
eigenstate of the operator ${S}_{z}$ of spin projection on the
axis $z$ is:
\begin{equation}
\label{hatn3} \hat {N} = 1 + \frac{{2\pi \rho} }{{k^{2}}}\left( {d
+ d_{1} m_{}^{2} }  \right),
\end{equation}
where $m$ is the magnetic quantum number.

\begin{figure}[!h]
\includegraphics[width=8cm,keepaspectratio]{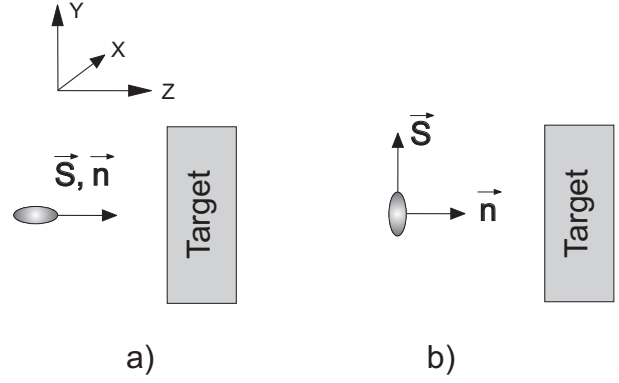}
\caption{Two possible orientation of vectors $\vec {S}$ and $\vec
{n}=\frac{\vec{k}}{k}$: a) $\vec {S} {\|} \vec {n}$; b) $\vec {S}
{\bot} \vec {n}$}.
 \label{sigma}
\end{figure}

According to Eq.(\ref{hatn3}), the refractive indices for the
states with $m=+1$ and $m=-1$ are the same, while those for $m=\pm
1$ and $m=0$ are different ($\Re{\textit{N}(\pm 1)} \neq
\Re{\textit{N}(0)}$ and $\Im{\textit{N}(\pm 1)} \neq
\Im{\textit{N}(0)}$).
This can be obviously explained as follows (see Fig.\ref{sigma}):
the shape of a deuteron in the ground state is non--spherical.
Therefore, the scattering cross-section $\sigma_{\|}$ for a
deuteron with $m= \pm 1$ (deuteron spin is parallel to its
momentum $\vec{k}$) differs from the scattering cross-section
$\sigma_\bot$ for a deuteron with $m=0$:
\begin{equation}
\sigma_{\|} \ne \sigma_\bot ~\Rightarrow ~
 \label{eq7} \Im\textit{f}_\|(0)=\frac{k}{4\pi}
\sigma_{\|}\neq \Im\textit{f}_ \bot (0)=\frac{k}{4\pi}\sigma_\bot.
\end{equation}
According to the dispersion relation $\Re{\textit{f}(0)} \sim\Phi
(\Im{\textit{f}(0))}$, then $\Re{\textit{f}_{ \bot }(0)} \neq
\Re{\textit{f}_{\|}(0)}$.

From the above it follows that deuteron spin dichroism appears
even when a deuteron passes through an unpolarized target: due to
different absorption the initially unpolarized beam acquires
polarization or, yet more precisely, alignment.

%%%%%%%%%%%%%%%%%%%%%%%

Let us consider the behavior of the deuteron spin in a target.
The spin state of the deuteron is described by its vector and
tensor
 polarization ($\vec {p} = \langle \vec {S}\rangle$ and $p_{ik} =
 \langle Q_{ik} \rangle$, respectively).
When the deuteron moves in matter its vector and tensor
polarization appears changed.
To calculate $\vec {p}$ and $p_{ik}$ one need to know the explicit
form of the deuteron spin wave function $\psi$

The wave function of the deuteron that has passed the distance $z$
inside the target is:
\begin{equation}
\label{psiz} \psi \left( {z} \right) = \exp\left( {ik\hat{N}z}
\right)\psi _{0},
\end{equation}
where $\psi_{0}$ is the wave function of the deuteron before
entering the target.
The wave function $\psi$ can be expressed as a superposition of
the basic spin functions $\chi_{m}$, which are the eigenfunctions
of the operators $\hat{S}^{2}$ and $\hat{S}_{z}$ ($\hat {S}_{z}
\chi _{m} = m\chi _{m}$):
\begin{equation}
\label{psi} \psi = \sum\limits_{m = \pm 1,0} {a^{m}\chi _{m}}  .
\end{equation}
Therefore,
\begin{equation}
 \label{psidepth}
\begin{array}{l}
 \Psi = \left( \begin{array}{*{20}c}
 {a^{1}} \hfill \\
 {a^{0}} \hfill \\
 {a^{ - 1}} \hfill \\
\end{array}  \right) = \left( {{\begin{array}{*{20}c}
 {ae^{i\delta _{1}} e^{ikN_{1} z}} \hfill \\
 {be^{i\delta _{0}} e^{ikN_{0} z}} \hfill \\
 {ce^{i\delta _{ - 1}} e^{ikN_{ - 1} z}} \hfill \\
\end{array}} } \right) = \\ \\ \qquad {}\left( {{\begin{array}{*{20}c}
 {ae^{i\delta _{1}} e^{ikN_{1} z}} \hfill \\
 {be^{i\delta _{0}} e^{ikN_{0} z}} \hfill \\
 {ce^{i\delta _{ - 1}} e^{ikN_{1} z}} \hfill \\
\end{array}}} \right),
\end{array}
\end{equation}
(according to the above $N_{1}=N_{-1}$).

Suppose the plane $(yz)$ coincides with the plane formed by the
initial deuteron vector polarization $\vec {p}_0 \neq 0$ and the
momentum $\vec{k}$ of the deuteron. In this case $\delta
_{1}-\delta _{0}= \delta _{0}-\delta _{-1}=\frac{{\pi}}{{2}}$, and
the components of the polarization vector at $z = 0$ are $p_x =
0,p_y\neq 0, \mbox{and }p_z\neq 0$.

The components of the vector polarization $\vec{p}=\langle \vec
{S}\rangle = \frac{{\langle \Psi \left| {\vec {S}} \right|\Psi
\rangle} }{{\left\langle {{\Psi} } \mathrel{\left| {\vphantom
{{\Psi} {\Psi} }} \right. \kern-\nulldelimiterspace} {{\Psi} }
\right\rangle} }$ inside the target are:
\\
\begin{eqnarray}
p_x&=& \frac{{ \sqrt {2} e^{ - \frac{{1}}{{2}}\rho \left( {\sigma
_{0} + \sigma _{1}} \right)z}b\left( {a - c} \right)\sin\left(
{\frac{{2\pi \rho} }{{k}}\Re d_{1} z} \right)}}{{\left\langle
{{\Psi} } \mathrel{\left| {\vphantom {{\Psi}  {\Psi} }} \right.
\kern-\nulldelimiterspace}
{{\Psi} } \right\rangle} } ,\nonumber\\
p_y&=&\frac{{\sqrt {2} e^{ -
\frac{{1}}{{2}}\rho \left( {\sigma _{0} + \sigma _{1}}
\right)z}b\left( {a + c} \right)\cos\left( {\frac{{2\pi \rho}
}{{k}}\Re d_{1} z} \right)}}{{\left\langle {{\Psi} } \mathrel{\left|
{\vphantom {{\Psi}  {\Psi} }} \right. \kern-\nulldelimiterspace}
{{\Psi} } \right\rangle} } ,\nonumber \\
p_z& =&\frac{{e^{\rho \sigma
_{1} z}\left( {a^{2} - c^{2}} \right)}}{{\left\langle {{\Psi} }
\mathrel{\left| {\vphantom {{\Psi}  {\Psi }}} \right.
\kern-\nulldelimiterspace} {{\Psi} } \right\rangle} }.\label{rot1}
\\ \nonumber
\end{eqnarray}
Similarly, the components of the tensor polarization $\hat
{Q}_{ij} = \frac{{3}}{{2}}\left( {\hat {S}_{i} \hat {S}_{j} + \hat
{S}_{j} \hat {S}_{i} - \frac{{4}}{{3}}\delta _{ij}} \right)$ are
expressed as:
\\
\begin{eqnarray}
p_{xx}&=&\frac{{ - \frac{{1}}{{2}}\left( {a^{2} +
c^{2}} \right)e^{ - \rho \sigma _{1} z} + b^{2}e^{ - \rho \sigma
_{0} z} - 3ace^{ - \rho \sigma _{1} z}}}{{\left\langle {{\Psi} }
\mathrel{\left| {\vphantom {{\Psi}  {\Psi} }} \right.
\kern-\nulldelimiterspace} {{\Psi} } \right\rangle} }  ,\nonumber
\\
p_{yy}&=&\frac{{ - \frac{{1}}{{2}}\left( {a^{2}
+ c^{2}} \right)e^{ - \rho \sigma _{1} z} + b^{2}e^{ - \rho \sigma
_{0} z} + 3ace^{ - \rho \sigma _{1} z}}}{{\left\langle {{\Psi} }
\mathrel{\left| {\vphantom {{\Psi}  {\Psi} }} \right.
\kern-\nulldelimiterspace} {{\Psi} } \right\rangle} }  , \nonumber
\\
p_{zz}&=&\frac{{\left( {a^{2} + c^{2}}
\right)e^{ - \rho \sigma _{1} z} - 2b^{2}e^{ - \rho \sigma _{0}
z}}}{{\left\langle {{\Psi} } \mathrel{\left| {\vphantom {{\Psi}
{\Psi} }} \right. \kern-\nulldelimiterspace} {{\Psi} }
\right\rangle} } ,\nonumber\\
p_{xy}&=& 0 ,\nonumber
\\
p_{xz}&=&\frac{{ \frac{{3}}{{\sqrt {2}} }e^{ - \frac{{1}}{{2}}\rho
\left( {\sigma _{0} + \sigma _{1}} \right)z}b\left( {a + c}
\right)\sin\left( {\frac{{2\pi \rho} }{{k}}\Re d_{1} z}
\right)}}{{\left\langle {{\Psi} } \mathrel{\left| {\vphantom
{{\Psi}  {\Psi }}} \right. \kern-\nulldelimiterspace} {{\Psi} }
\right\rangle} }  ,\nonumber
\\
p_{yz}&=&\frac{{\frac{{3}}{{\sqrt
{2}} }e^{ - \frac{{1}}{{2}}\rho \left( {\sigma _{0} + \sigma _{1}}
\right)z}b\left( {a - c} \right)\cos\left( {\frac{{2\pi \rho}
}{{k}}\Re d_{1} z} \right)}}{{\left\langle {{\Psi} } \mathrel{\left|
{\vphantom {{\Psi}  {\Psi }}} \right. \kern-\nulldelimiterspace}
{{\Psi} } \right\rangle} } ,
\label{rot2}
\end{eqnarray}
\\
\noindent where $\left\langle {{\Psi} } \mathrel{\left| {\vphantom {{\Psi}
{\Psi} }} \right. \kern-\nulldelimiterspace} {{\Psi} }
\right\rangle = \left( {a^{2} + c^{2}} \right)e^{ - \rho \sigma
_{1} z} + b^{2}e^{ - \rho \sigma _{0} z}$,
$\sigma_{0}=\frac{{4\pi} }{{k}}\Im f_0$, $\sigma _{1} = \frac{{4\pi}
}{{k}}\Im f_1$, $f_0=d$, $f_1=d+d_1$.

According to (\ref{rot1},\ref{rot2}) the spin rotation occurs when
the angle between the polarization vector $\vec{p}$ and momentum
$\vec{k}$ of the particle differs from $\frac{{\pi} }{{2}}$.

For example, when $\Re d_1 >0$, the angle between the polarization
vector and momentum is acute and the spin rotates left-hand around
the momentum direction, whereas the obtuse angle between the
polarization vector and momentum gives rise to the right-hand spin
rotation.

When the polarization vector and momentum are perpendicular
 (transversely polarized particle), then  the components of
 the vector polarization at $z = 0$ are: $p_x= 0$, $p_y\ne0$, and $p_z=0$.
In this case $a=c$ and the dependance of the vector polarization
on $z$ can be expressed as:
\begin{eqnarray}
p_x&=&0,\nonumber\\
p_y&=&\frac{{\sqrt {2} e^{ -
\frac{{1}}{{2}}\rho \left( {\sigma _{0} + \sigma _{1}}
\right)z}2ba\cos\left( {\frac{{2\pi \rho }}{{k}}\Re d_{1} z}
\right)}}{{\left\langle {{\Psi} } \mathrel{\left| {\vphantom
{{\Psi}  {\Psi} }} \right.
\kern-\nulldelimiterspace} {{\Psi} } \right\rangle} },\nonumber\\
p_z&=&0,\nonumber\\
p_{xx}&=&\frac{{ - 4a^{2}e^{ - \rho \sigma _{1} z}
+ b^{2}e^{ - \rho \sigma _{0} z}}}{{\left\langle {{\Psi} }
\mathrel{\left| {\vphantom {{\Psi}  {\Psi} }} \right.
\kern-\nulldelimiterspace} {{\Psi} } \right\rangle} }, \\
p_{yy}&=&\frac{{2a^{2}e^{ - \rho \sigma _{1} z} +
b^{2}e^{ - \rho \sigma _{0} z}}}{{\left\langle {{\Psi} }
\mathrel{\left| {\vphantom {{\Psi}  {\Psi} }} \right.
\kern-\nulldelimiterspace} {{\Psi} } \right\rangle} },\nonumber\\
p_{zz}&=&\frac{{2a^{2}e^{ - \rho \sigma_{1} z} -
2b^{2}e^{ - \rho \sigma _{0} z}}}{{\left\langle {{\Psi} }
\mathrel{\left| {\vphantom {{\Psi}  {\Psi} }} \right.
\kern-\nulldelimiterspace} {{\Psi} } \right\rangle} }, \nonumber
\\
p_{xz}&=&\frac{{  \frac{{3}}{{\sqrt {2}} }e^{ -
\frac{{1}}{{2}}\rho \left( {\sigma _{0} + \sigma _{1}}
\right)z}2ab\sin\left( {\frac{{2\pi \rho} }{{k}}\Re d_{1} z}
\right)}}{{\left\langle {{\Psi} } \mathrel{\left| {\vphantom
{{\Psi}  {\Psi }}} \right. \kern-\nulldelimiterspace} {{\Psi} }
\right\rangle} },\nonumber\\
p_{yz}&=&0. \nonumber \label{11}
\end{eqnarray}
According to (\ref{11}) the vector and tensor polarization
oscillate when a transversely polarized deuteron passes through
matter.

\section{SPIN ROTATION AND OSCILLATION AND SPIN DICHROISM
OF A 20 MeV STORED DEUTERON BEAM}

Let us consider the 20 MeV deuteron beam passing through
unpolarized matter (this energy is typical for low-energy
accelerators).
%
%-----------------------------
{The density matrix for the deuteron beam before the target can be
written as follows: }
 \begin{eqnarray}
\hat{\rho}_{0} &=& \frac{{\hat {I}}}{{3}} + \frac{{1}}{{2}}\left(
{{\vec{p}_{x}}\hat{\vec{S}_{x}}+{\vec{p}_{y}}\hat {\vec{S}_{y}}+
{\vec{p}_{z}}\hat{\vec{S}_{z}}}  \right)\nonumber\\
&+& \frac{{2}}{{9}}\left( {{p_{xy}}\hat {Q}_{xy} +{p_{xz}}\hat{Q}_{xz}
 + {p_{yz}} \hat {Q}_{yz}}\right)\nonumber\\
&+&  \frac{{1}}{{9}}\left( {{p_{xx}}\hat{Q}_{xx}
+{p}_{yy}}\hat{Q}_{yy} + {p_{zz}}\hat{Q}_{zz}\right).\label{rhoexp}\\
\nonumber
\end{eqnarray}
Using (\ref{psidepth}) we can express the density matrix of the
deuteron beam in the target as:
\begin{equation}
\begin{array}{l}
 \hat {\rho}  = \left( {{\begin{array}{*{20}c}
 {e^{ikzN_{1}} } \hfill & {0} \hfill & {0} \hfill \\
 {0} \hfill & {e^{ikN_{0} z}} \hfill & {0} \hfill \\
 {0} \hfill & {0} \hfill & {e^{ikN_{1} z}} \hfill \\
\end{array}} } \right)\hat {\rho} _{0} \times \\\\ \qquad
\left( {{\begin{array}{*{20}c}
 {e^{ - ikN_{1}^{\ast}  z}} \hfill & {0} \hfill & {0} \hfill \\
 {0} \hfill & {e^{ - ikN_{0}^{\ast}  z}} \hfill & {0} \hfill \\
 {0} \hfill & {0} \hfill & {e^{ - ikN_{1}^{\ast}  z}} \hfill \\
\end{array}} } \right)
\end{array}
\end{equation}
then
\[
\vec{p}=\langle \vec{S}\rangle = \frac{{{\rm{Tr}} \left( \hat {\rho} \hat
{\vec{ S}}\right)}}{{ \rm{Tr}\left(\hat {\rho} \right)}},\quad p_{ik}= \langle
Q_{ik} \rangle = \frac{{\rm{Tr}} \left( \hat {\rho}\hat{Q}_{ik}
\right)}{ \rm{Tr}\left(\hat {\rho} \right)}
\]where $i,k=x,y,z$.
Suppose $p_{x,0}$, $p_{y,0}$, $p_{z,0}$, $p_{xx,0}$, $p_{yy,0}$,
$p_{zz,0}$, $p_{xy,0}$, $p_{xz,0}$,
 $p_{yz,0}$
 are the components of the deuteron vector and tensor polarization before entering the target then
the vector and tensor polarization of the deuteron inside the
target can be expressed using the first-order approximation
$e^{ikz\left( {N_{1} - N_{1}^{\ast} } \right)}\approx1+ikz\left(
{N_{1} - N_{1}^{\ast} } \right)$ as:
\begin{eqnarray}
p_{x}&=& \frac{{\left[ {1 - \frac{{1}}{{2}}\rho z\left( {\sigma
_{0} + \sigma _{1}}  \right)} \right] p_{x,0}+
\frac{{4}}{{3}}\frac{{\pi \rho z}}{k}\Re d_{1} p_{zy,0}}}{{Tr\hat
{\rho }\hat {I}}},\nonumber
\\
p_{y} &=& \frac{{\left[ {1 - \frac{{1}}{{2}}\rho z\left( {\sigma
_{0} + \sigma _{1}}  \right)} \right] p_{y,0}  -
\frac{{4}}{{3}}\frac{{\pi \rho z}}{k} \Re d_{1} p_{zx,0}}}{{Tr\hat
{\rho }\hat {I}}}, \nonumber
\\
p_{z}& =&
\frac{{\left( {1 - \rho \sigma _{1} z} \right)p_{z,0}} }
{{Tr\hat {\rho} \hat {I}}},\nonumber
 \\ \nonumber \\
p_{xx}&=&{\textstyle  \frac{{\left( {1 - \rho \sigma _{1} z}
\right)p_{xx,0} + \frac{{1}}{{3}}\rho z\left( {\sigma _{1} -
\sigma _{0}} \right) - \frac{{1}}{{3}}\rho z\left( {\sigma _{1} -
\sigma _{0}}
\right)p_{zz,0}} }{{Tr\hat {\rho} \hat {I}}},} \nonumber \\
p_{yy} &=&{\textstyle \frac{{\left( {1 - \rho \sigma _{1} z}
\right)p_{yy,0} + \frac{{1}}{{3}}\rho z\left( {\sigma _{1} -
\sigma _{0}}  \right) - \frac{{1}}{{3}}\rho z\left( {\sigma _{1} -
\sigma _{0}} \right)p_{zz,0}}
}{{Tr\hat {\rho} \hat {I}}},} \nonumber\\
p_{zz} &=& \frac{{\left[ {1 - \frac{{1}}{{3}}\rho z\left( {2\sigma
_{0} + \sigma _{1}}  \right)} \right]p_{zz,0} -
\frac{{2}}{{3}}\rho z\left( {\sigma _{1} - \sigma _{0}}
\right)}}{{Tr\hat {\rho} \hat {I}}}, \nonumber\\
p_{xy} &=&
\frac{{\left( {1 - \rho \sigma _{1} z} \right)p_{xy,0}} }
{{Tr\hat {\rho} \hat {I}}},\label{polpar1}
\\p_{xz}
&=& \frac{{\left[ {1 - \frac{{1}}{{2}}\rho z\left( {\sigma _{0} +
\sigma _{1}}  \right)} \right] p_{xz,0} + 3\frac{{\pi \rho
z}}{{k}}\Re d_{1} p_{y,0} } }{{Tr\hat
{\rho }\hat {I}}},\nonumber \\
p_{yz} &=& \frac{{\left[ {1 - \frac{{1}}{{2}}\rho z\left( {\sigma
_{0} + \sigma _{1}} \right)} \right]p_{yz,0} - 3\frac{{\pi \rho
z}}{{k}}\Re d_{1} p_{x,0}} }{{Tr\hat {\rho }\hat {I}}}, \nonumber
\\ \nonumber
\end{eqnarray}
\noindent where $Tr\hat {\rho} \hat {I} = 1 - \frac{{\rho
z}}{{3}}\left( {2\sigma _{1} + \sigma _{0}}  \right) - \frac{{\rho
z}}{{3}}\left( {\sigma _{1} - \sigma _{0}} \right)p_{zz,0} .$

If the beam is initially unpolarized ($p_{x,0} = p_{y,0} = p_{z,0}
= p_{xx,0} = p_{yy,0} =$ $p_{zz,0} = p_{xy,0} = p_{xz,0} =
p_{yz,0} = 0$) then after passing through the unpolarized target
of thickness $z$ the deuteron beam acquires the tensor
polarization:
\begin{eqnarray}
p_{zz}&\approx& - \frac{{2}}{{3}}\rho z\left( {\sigma _{1} -
\sigma _{0}}  \right),\nonumber\\
p_{xx} &=& p_{yy} \approx \frac{{1}}{{3}}\rho z\left( {\sigma _{1}
- \sigma _{0}}  \right). \label{polpar2}
\end{eqnarray}
The tensor polarization can be expressed by dichroism $\textrm{G}$
of the unpolarized target:
\begin{equation}
p_{zz} \approx - \frac{{4}}{{3}}\textrm{G}, \quad p_{xx} = p_{yy}
\approx \frac{{2}}{{3}}\textrm{G} ,
\end{equation}
where
\begin{equation}
\textrm{G} = \frac{{I_{0} - I_{ \pm} } }{{I_{0} + I_{ \pm} } } =
\frac{{\rho z}}{{2}}\left( {\sigma _{1} - \sigma _{0}}  \right),
\end{equation}
%
% ---------------------------
${I}_{0}$ is the intensity of the deuteron beam after the target
if the deuteron beam before the target is in the spin state
${m}=0$ and, similarly, ${I}_{\pm} $ is the intensity of the
deuteron beam after the target if the deuteron beam before the
target is in the spin state $m=\pm 1$.
%-----------------------------------------------------

 Let us evaluate the angle of spin rotation and dichroism
of a 20 MeV deuteron beam in an unpolarized carbon target using
the above formulas: suppose the target density $\rho\approx
10^{23}$ cm$^{-3}$, the target thickness $z\approx$ 0.1 cm,
$\Re{d}_{1}\approx 6 \cdot10^{-13}$ cm,
$\sigma_{1}-\sigma_{0}$$\approx-10^{-25}$ cm$^{2}$ then the angle
of rotation (the magnitude of oscillation phase) is
$\varphi={\frac{{2\pi \rho} }{{k}}\Re d_{1} z} \sim 10^{-3}$ rad
and dichroism is $\textrm{G} \sim 10^{-2}$.

The above evaluation considers only the non-spherical shape of a
deuteron in the ground state, however an additional correction may
occur when considering the spin-spin and Coulomb interactions at
low energies.
Moreover, the eikonal formalism \cite{bar99} can not be applied to
the low energy deuterons because the wavelength of a 20 MeV
deuteron is compared with the nucleus size.
Nevertheless, when suppose that the S-wave contributes the most to
the scattering of a deuteron with the energy $<$ 20 MeV by a
nucleus, then consideration shows that the difference
$(\sigma_1-\sigma_0)$ (and hence, dichroism $\textrm{G}$) has the
same magnitude, while the sign is opposite.

\section{THE EXPERIMENT FOR DETECTION OF DEUTERON SPIN DICHROISM WITH A
{\bf $^{3}$He}~-~POLARIMETER}

The experiment for the detection of spin dichroism of deuterons in
a carbon target was carried out at the accelerator of the
Institute of Nuclear Physics of Cologne University. The existing
$^{3}$He polarimeter of the experimental installation was used.
The purpose was to measure all components of the deuteron vector
and tensor polarization via anisotropies of the protons outgoing
from the nuclear reaction \cite{ohl72,eng97}
\begin{equation}
 d +{}^{3}He \to {}^{4}He + p ~.
\end{equation}
The polarimeter has four detectors at $24.5^{ \circ}$ polar angles
in addition to the fifth detector measuring the emitted protons in
the forward direction (at $0^{ \circ}$).
This detector is sensitive only to the $p_{zz}$-component of the
tensor polarization \cite{eng97}.

Let us briefly describe the measurement procedure.
%
%% ----------------------------Ins(1)
The most general form for the cross section of a parity conserving
reaction induced by polarized spin 1 particle is \cite{ohl72}:
\begin{eqnarray} I(\theta ,\varphi) &=& I_{0}
(\theta) \left[ 1 + \frac{3}{2}p_{y} {A}_{y}(\theta) \right.
\nonumber\\ & + & \frac{2}{3}p_{xz}
A_{xz} (\theta) + \frac{1}{3}p_{xx} A_{xx} (\theta) \nonumber \\
&+&
 \left. \frac{1}{3}p_{yy} A_{yy}(\theta) + \frac{1}{3}p_{zz} A_{zz} (\theta) \right],
 \label{I_theta}
\end{eqnarray}
where  we use the projectile helicity coordinate system to
describe scattering or reactions induced by polarized spin 1
particles. In this coordinate system the axis of quantization
 ($z$ axis) is taken along the direction of the projectile motion
 $\vec{k}$; the $y$ axis is taken along $ \vec{k} \times
 \vec{k}^{\prime}$,
 where $\vec{k}^{\prime}$ represents the direction of scattered
 particle or reaction product motion; and the $x$ axis is chosen to
 form a right-handed coordinate system,
$\theta$ is the angle between $\vec{k}$ and $\vec{k}^{\prime}$,
and $A$ is the analyzing power of the reaction.
\begin{figure}[!h]
\includegraphics[width=8cm,keepaspectratio]{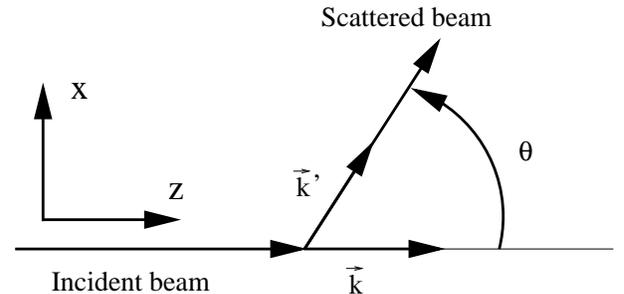}
\caption{The system of coordinates.} \label{sistem}
\end{figure}
The incident beam may have vector polarization components $p_x$,
$p_y$ and $p_z$, but because of parity conservation the reaction
is sensitive only to the vector component normal to the scattering
plane.
 Similarly, although the incident beam may contain all six tensor
 polarization components $p_{xy}$, $p_{yz}$, $p_{xz}$, $p_{xx}$,
 $p_{yy}$ and $p_{zz}$, the reaction is sensitive, again because
 of parity, only to those terms indicated in equation
 (\ref{I_theta}).
The diagonal elements of the tensor polarization meet the trace
relation:
\begin{equation}
p_{xx} + p_{yy} + p_{zz} = 0  .
\label{tracenull1}
\end{equation}
Similarly, for the analyzing power components:
\begin{equation}
\label{tracenull2} A_{xx} + A_{yy} + A_{zz} = 0~.
\end{equation}
Thus, only four of the five analyzing tensors that appear in
(\ref{I_theta}) are independent.

% ----------------replacement ---------------
%
The expression of equation (\ref{I_theta}) implies scattering to
the left, if $\vec{k} \times \vec{k^{\prime}}$ defines the 'up'
direction. If we define the azimuthal angle, $\varphi$, to be the
angle between the plane containing both $\vec{k} \times
\vec{k^{\prime}}$ and $\vec{k}$, and some fixed reference plane,
which also contains $\vec{k}$, the most general possible azimuthal
dependence for the cross section is:
\begin{eqnarray}
I(\theta ,\varphi) &=&  I_{0}(\theta)\times \nonumber\\ &\times
&\left\{ 1 + \frac{3}{2}({p_{x'} \sin\varphi + p_{y'}
\cos\varphi})A_{y} (\theta) \right.\nonumber\\ &+&
 \frac{2}{3}({p_{x'z'} \cos \varphi -
p_{y'z'} \sin \varphi})A_{xz}(\theta)\nonumber\\ &+&
\frac{1}{6}\left[ (p_{x'z'} - p_{y'y'})\cos2 \varphi - 2p_{x'y'}
\sin 2\varphi \right] \times \nonumber\\&\times &\left.
\left[A_{xx}(\theta) - A_{yy}(\theta)\right] + \frac{1}{2}p_{z'z'}
A_{zz} (\theta)\right\} \label{I_phi}
\end{eqnarray}
%---------------------------------------->>>>>>>>>

\begin{figure}[!h]
\includegraphics[width=8cm,keepaspectratio]{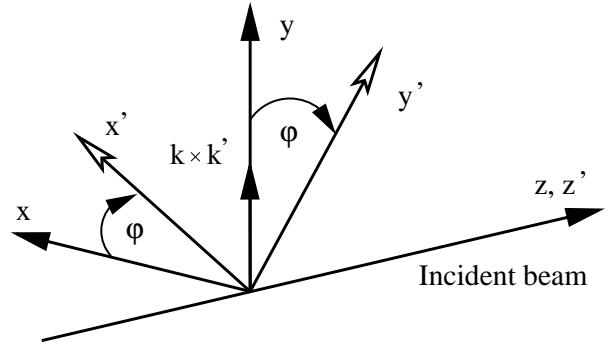}
\caption{Relation between the projectile helicity frame ($x$, $y$,
$z$) and the frame in terms of which the beam polarization is
described ($x^{\prime}$, $y^{\prime}$, $z^{\prime}$). For left
scattering ($\varphi=0$) these frames coincide.}
 \label{sistem2}
\end{figure}

%
%-------------definition of coordinate frame
The frame in terms of which the beam polarization is described
($x^{\prime}$, $y^{\prime}$, $z^{\prime}$) is defined according to
\cite{ohl72}: suppose the beam has components of the vector
polarization $p_{x^{\prime}}$, $p_{y^{\prime}}$, $p_{z^{\prime}}$,
where ${z^{\prime}}$ is along the beam direction and
${y^{\prime}}$ is chosen in a way that is natural for the
polarized beam.
 For example, if the beam is prepared by a polarized ion source,
 ${y^{\prime}}$ is chosen so that its polarization axis of
 symmetry lies in the (${y^{\prime}}$, ${z^{\prime}}$) plane with
 positive ${y^{\prime}}$; if it is prepared by scattering,
 ${y^{\prime}}$ is chosen along $\vec{k} \times \vec{k^{\prime}}$
 for scattering. Fig.\ref{sistem2} shows the relation between
 ($x^{\prime}$, $y^{\prime}$, $z^{\prime}$) (the beam coordinate system)
 and ($x$, $y$,
$z$) (the analyzer coordinate system).

Using (\ref{I_phi}) we can express the number of events registered
by the detectors of the $^{3}$He-polarimeter: $L$ corresponds to
the left detector ($\theta$~=~24.5$^{\circ}$,
$\varphi$~=~0$^\circ$), $R$, $U$, $D$ and $F$ are for the right
($\theta$~=~24.5$^\circ$, $\varphi$~=~180$^\circ$), top
($\theta$~=~24.5$^\circ$, $\varphi$~=~270$^\circ$), bottom
($\theta$~=~24.5$^\circ$, $\varphi$~=~90$^\circ$) and forward
($\theta$~=~0$^\circ$) detectors, respectively.
\begin{eqnarray*}
L &=& Nn\Omega _{L} E I_{0}(24.5^\circ)\times \\& \times &
\left\{1+\frac{3}{2}p_{y'} A_{y} (24.5^\circ) +
\frac{2}{3}p_x'z'A_{xz} (24.5^\circ)\right.\\ &+&
 \frac{1}{6}(p_{x'z'} - p_{y'y'})\left[A_{xx}(24.5^\circ) -
A_{yy}(24.5^\circ) \right] \\ &+&\left. \frac{1}{2}p_{z'z'}
A_{zz}(24.5^\circ)\right\};
\end{eqnarray*}
\begin{eqnarray*}
R &=& Nn\Omega _{R} E I_{0}(24.5^\circ)\times \\&\times&
\left\{1-\frac{3}{2}p_{y'} A_{y} (24.5^\circ) -
\frac{2}{3}p_x'z'A_{xz} (24.5^\circ)\right.\\ &+&
 \frac{1}{6}(p_{x'z'} - p_{y'y'})\left[A_{xx}(24.5^\circ) -
A_{yy}(24.5^\circ) \right] \\ &+&\left. \frac{1}{2}p_{z'z'}
A_{zz}(24.5^\circ)\right\};
\end{eqnarray*}
\begin{eqnarray*}
U &=& Nn\Omega _{U} E I_{0}(24.5^\circ)\times \\&\times&
\left\{1-\frac{3}{2}p_{y'} A_{y} (24.5^\circ) +
\frac{2}{3}p_x'z'A_{xz} (24.5^\circ)\right.\\ &-&
 \frac{1}{6}(p_{x'z'} - p_{y'y'})\left[A_{xx}(24.5^\circ) -
A_{yy}(24.5^\circ) \right] \\ &+&\left. \frac{1}{2}p_{z'z'}
A_{zz}(24.5^\circ)\right\};
\end{eqnarray*}
\begin{eqnarray*}
D &=& Nn\Omega _{D} E I_{0}(24.5^\circ)\times \\&\times&
\left\{1+\frac{3}{2}p_{y'} A_{y} (24.5^\circ) -
\frac{2}{3}p_x'z'A_{xz} (24.5^\circ)\right.\\ &-&
 \frac{1}{6}(p_{x'z'} - p_{y'y'})\left[A_{xx}(24.5^\circ) -
A_{yy}(24.5^\circ) \right] \\ &+&\left. \frac{1}{2}p_{z'z'}
A_{zz}(24.5^\circ)\right\};
\end{eqnarray*}
\begin{eqnarray*}
F &=& Nn\Omega _{F} E I_{0} (0^\circ)\left[1+\frac{1}{2}p_{z'z'}
A_{zz}(0^\circ) \right],
\end{eqnarray*}
where $N$ is the surface density of the $^{3}\rm{He}$ target in
cm$^{-2}$, $n$ is the number of incident deuterons, $\Omega$ is
the solid angle, and $E$ is the efficiency of each detector.

\begin{figure}[!h]
\includegraphics[width=8cm,keepaspectratio]{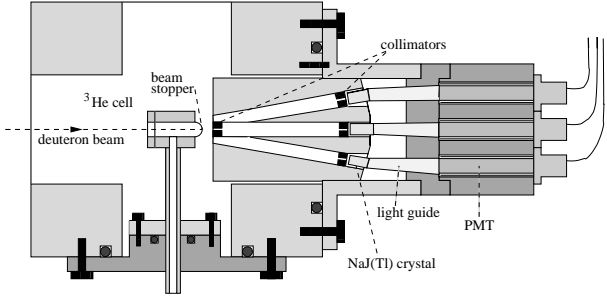}
\caption{$^{3}\rm{He}$-Polarimeter.}
 \label{polar}
\end{figure}

\begin{figure}[!h]
\includegraphics[width=8cm,keepaspectratio]{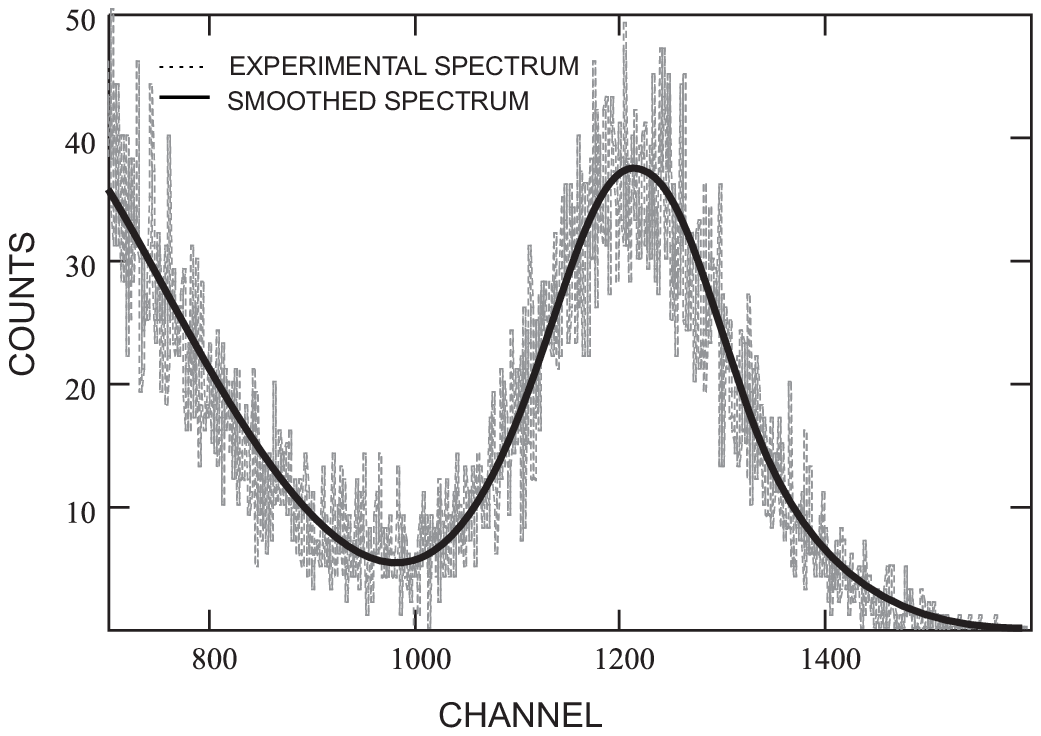}
\caption{Energy spectrum of protons registered by one side
detector, produced by deuterons with an initial energy of 16.2 MeV
after passing through the 151~mg/cm$^2$ carbon target.}
 \label{rightspec}
\end{figure}

\begin{figure}[!h]
\includegraphics[width=8cm,keepaspectratio]{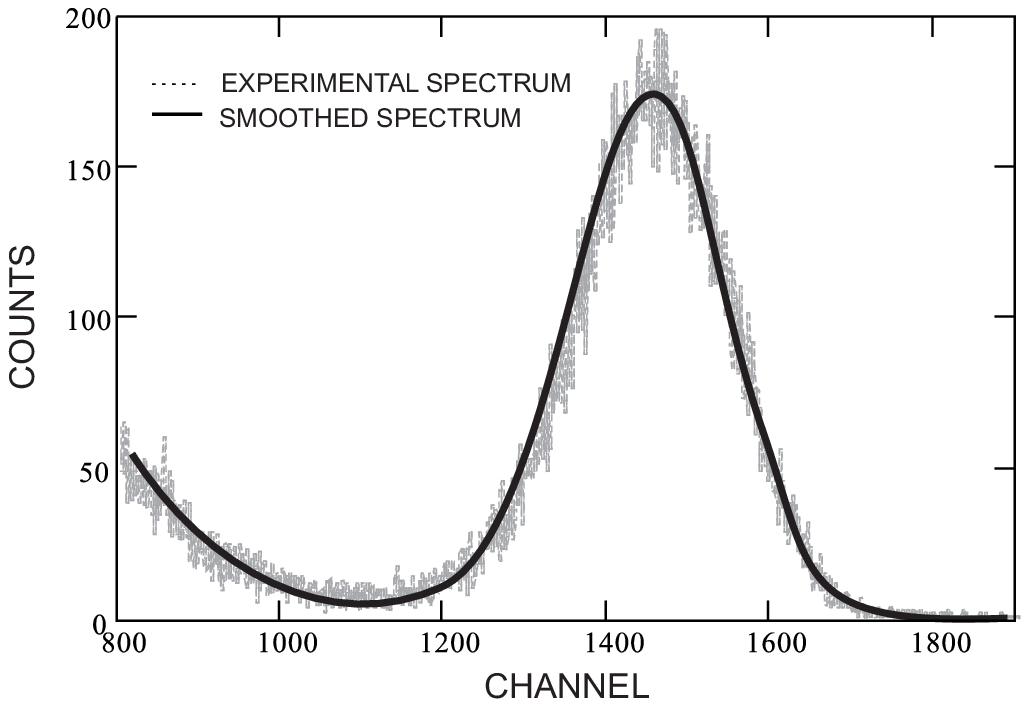}
\caption{Energy spectrum of protons registered by the forward
detector, produced by deuterons with an initial energy of 16.2 MeV
after passing through the 151~mg/cm$^2$ carbon target.}
 \label{forspec}
\end{figure}

\begin{figure}[!h]
\includegraphics[width=9cm,keepaspectratio]{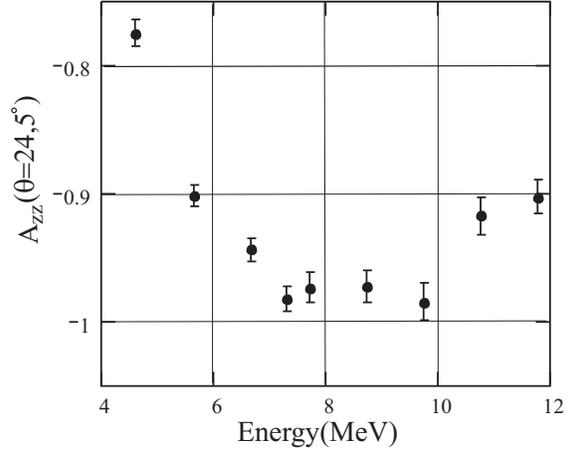}
\caption{Analyzing power for the side detectors.}
 \label{azz245}
\end{figure}

\begin{figure}[!h]
\includegraphics[width=9cm,keepaspectratio]{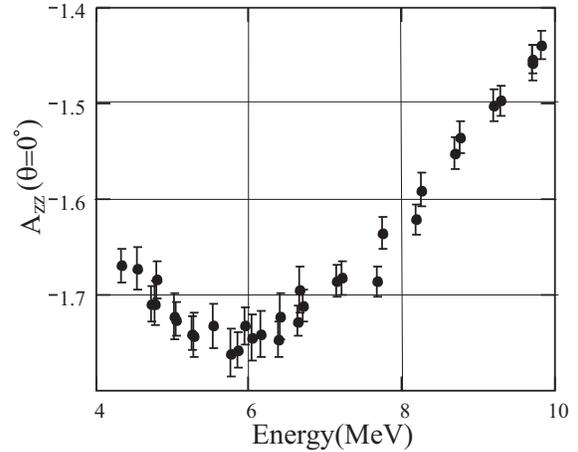}
\caption{Analyzing power for the forward detector.}
 \label{azz0}
\end{figure}

The scheme of the polarimeter \cite{eng97} is shown in
Fig.\ref{polar}, Figs. \ref{rightspec}-\ref{azz0} display typical
spectrums of detected protons produced by deuterons in the
$^{3}\rm{He}$ cell and analyzing power for the side and forward
detectors.

\section{DISCUSSION: GENERAL APPROACH TO SPIN DICHROISM MEASUREMENT IN VIEW OF
{\bf $^{3}$He}-POLARIMETER FEATURES}

Dichroism and tensor polarization of the deuteron beam passing
through the carbon target was evaluated in the previous sections
as $\sim $10$^{-2}$.
Therefore, the number of particles detected when deuteron beam
passed through the target is two orders less than the number of
particles detected without a target.

During the experiment several issues arose.
The first is associated with the dependence of the detector count
rate on beam focusing.

{The beam crossover with no target was small and it could cross
the helium cell in different places resulting in the variation in
solid angle ratios between different detectors.}
Thus, individual numbers of  events registered by each detector
could vary in about 5 \%.

Incomplete suppression of secondary electrons in the polarimeter
caused the second issue: the number of counts for the equal charge
also substantially depended on focusing.

\begin{figure}[!h]
\includegraphics[width=8cm,keepaspectratio]{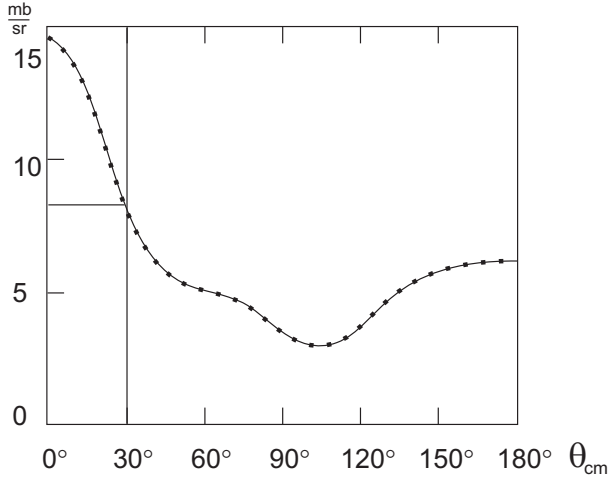}
\caption{Differential cross section of the
$^{3}\rm{He}(d,p)^{4}\rm{He}$ reaction at 6~MeV.}
 \label{cros6}
\end{figure}

\begin{figure}[!h]
\includegraphics[width=8cm,keepaspectratio]{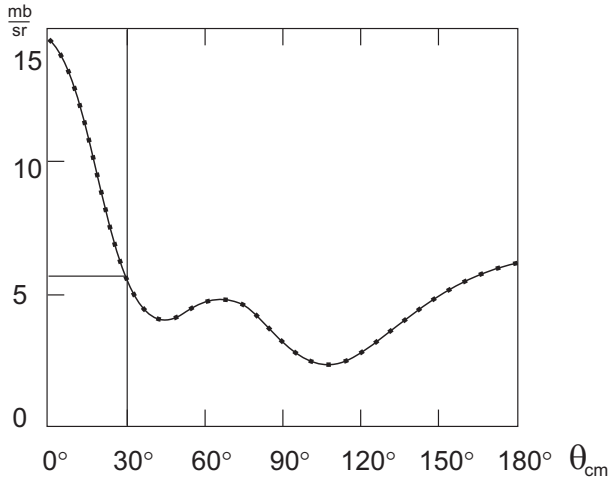}
\caption{Differential cross section of the
$^{3}\rm{He}(d,p)^{4}\rm{He}$ reaction at 8~MeV.}
 \label{cros8}
\end{figure}

The differential cross section of the reaction $d + {}^{3}He \to
{}^{4}He + p$ strongly depends on energy as it is shown in Fig.
\ref{cros6},\ref{cros8} \cite{Bittcher}.
For these reasons, it was impossible to calibrate the detectors
accurately.
Besides, the tensor polarization could be found only
approximately.
%------------------------------------------------------------------------------------------------
In the following experiment the measurement will be improved.
 It is necessary to choose such combinations of \textit{L},
\textit{R}, \textit{U}, \textit{D}, \textit{F} with and without a
target so that the results can become maximally insensitive to
solid-angle variations. Let us designate by $L_{t}$, $R_{t}$,
$U_{t}$, $D_{t}$, $F_{t}$ the number of the registered particles
with a target, and \textit{L}, \textit{R}, \textit{U}, \textit{D},
\textit{F} without. According to Eqs.
(\ref{polpar1}),(\ref{polpar2}), after the target the beam has
only diagonal components of the tensor polarization
\textit{p}$_{zz}$, \textit{p}$_{xx}$, \textit{p}$_{yy}$, but as
$p_{xx}=p_{yy}$, we obtain:

\begin{eqnarray*}
L_{t}&=& Nn_{t} \Omega_{L}EI_{0}(24.5^\circ)\left[1
+\textstyle{\frac{1}{2}}p_{z'z'} A_{zz}(24.5^\circ) \right]\\
L &=& Nn\Omega_{L} EI_{0}(24.5^\circ)\\
R_{t}&=& Nn_{t} \Omega _{R} EI_{0}(24.5^\circ)
\left[1 + \textstyle{\frac{1}{2}}p_{z'z'} A_{zz}(24.5^\circ) \right])\\
R &=& Nn\Omega _{R} EI_{0}(24.5^\circ)\\
U_{t} & =&Nn_{t} \Omega _{U} EI_{0}(24.5^\circ) \left[1 +
\textstyle{\frac{1}{2}}p_{z'z'} A_{zz}(24.5^\circ)
\right]\\
U &=& Nn\Omega _{U} EI_{0}(24.5^\circ)\\
D_{t} &=&Nn_{t} \Omega _{D} EI_{0}(24.5^\circ) \left[1 + \textstyle{\frac{1}{2}}p_{z'z'} A_{zz}(24.5^\circ) \right]\\
D &=& Nn\Omega _{D} EI_{0}(24.5^\circ)\\
F_{t} &=& Nn_{t} \Omega _{F} EI_{0}(0^\circ)
\left[1 + \textstyle{\frac{1}{2}}p_{z'z'} A'_{zz}(0^\circ)\right]\\
F &=& Nn\Omega _{F} EI_{0}(0^\circ)\\
\end{eqnarray*}

The relevant combinations used in the data reduction -- their variation
in this case was about 0.2\% -- were therefore:

\begin{equation} \frac{L + R + U + D}{F}\\ =
\frac{( \Omega _{L} + \Omega _{R} + \Omega _{U} + \Omega
_{D})I_{0}(24.5^\circ) }{\Omega _{F}I_{0}(0^\circ) }
\end{equation}

\begin{eqnarray}
&{ }&\frac{L_{t} + R_{t} + U_{t} + D_{t}} {F_{t}}\nonumber\\ &=&
\frac{(\Omega _{L} + \Omega _{R} + \Omega _{U} + \Omega _{D}
)I_{0}(24.5^\circ)}{\Omega _{F} I_{0}( 0^\circ)}\times \nonumber\\
&\times& \frac{ \left[ 1 + \frac{1}{2}p_{zz} A_{zz}(24.5^\circ)
\right]}{1 + \frac{1}{2}p_{zz} A'_{zz}(0^\circ)}
\end{eqnarray}

With $p_{zz}\ll$1, $A_{zz}\sim $1
\begin{eqnarray}
& &\frac{L_{t} + R_{t} + U_{t} + D_{t} }{F_{t}}\cong \frac{L + R +
U + D}{F} \times \nonumber \\ &\times& \left\{1 +
\textstyle{\frac{1}{2}}p_{zz}\left[A_{zz}(24.5^\circ) - A'_{zz}
(0^\circ) \right] \right\} \label{apline}
\end{eqnarray}

\noindent i.e., if the beam acquires the tensor polarization in
the target this should lead to a change in the ratio of the sum of
the counts number of the side detectors to the counts of the
forward detector.

The resulting energy dependence of the experimental points was
approximated by a linear least-squares fit. The measurements were
carried out with three different targets and the corresponding
energies of the incident deuteron beam. The deuteron energies
after the targets were found from the Bethe-Bloch formula and
should be about 7 MeV. For example, a deuteron beam of 18.1 MeV in
front of the carbon target with 188 $\frac {\footnotesize
\mbox{mg}}{\footnotesize \mbox{cm}^2}$ produces 7 MeV average beam
energy after the target. At this energy of the incident beam
calibration measurements without a target were taken before.

For each target a similar processing of the spectrum was done. The
thickness of each target and parameters of the linear fit for each case are
shown in table~\ref{tab1}.

\begin{table}[!h]
\caption{Thickness of targets and parameters of the linear fit.}
\begin{center}

\begin{tabular}{c|c|c}
 Target&\multicolumn{2}{c}{Parameters of}\\
thickness & \multicolumn{2}{c}{linear fit}\\
(mg/cm$^2$)&\multicolumn{2}{c}{y = kx + b}\\
\hline
&\textit{k}&
\textit{b} \\
\hline
0&
-0.134 \par $\pm $0.005&
1.72 \par $\pm $0.03 \\
\hline 57.8 \par $\pm $1.0&
 -0.128 \par $\pm $0.005&
1.68 \par $\pm $0.03 \\
\hline 151 \par $\pm $3&
 -0.103 \par $\pm $0.006&
1.50 \par $\pm $0.04 \\
\hline 188 \par $\pm $4&
-0.116 \par $\pm $0.005&
1.59 \par $\pm $0.03 \\
%%%%\hline
\end{tabular}
\end{center}
\label{tab1}
\end{table}

\begin{figure}[!h]
\includegraphics[width=9cm,keepaspectratio]{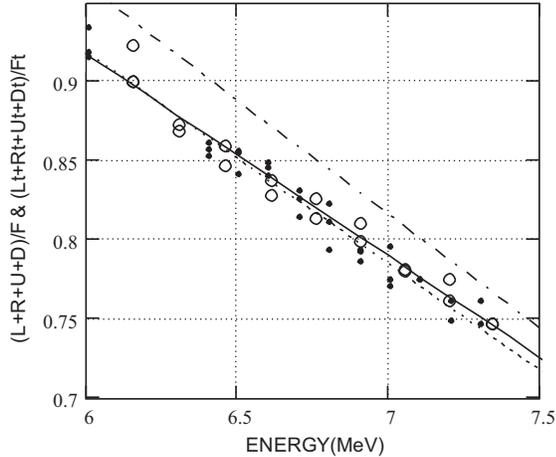}
\caption{Energy dependence of (L+R+U+D)/F without target and
(L$_{t}$+R$_{t}$+U$_{t}$+D$_{t}$)/F$_{t}$ for the target of 58
mg/cm$^{2}$ together with the corresponding linear approximations
( $\bullet $ \hspace{0.2cm} experimental points (L+R+U+D)/F
without a target (calibration) / $\circ $ \hspace{0.2cm}
experimental points (L+R+U+D)/F for the targets of 58 mg/cm$^{2}$,
151 mg/cm$^{2}$, 188 mg/cm$^{2}$, respectively / $\cdot \cdot
\cdot \cdot \cdot $ \hspace{0.2cm} the linear fit to experimental
points (L+R+U+D)/F without target /----------- \hspace{0.2cm} the
linear fit to experimental points (L+R+U+D)/F for the targets of
58 mg/cm$^{2}$, 151 mg/cm$^{2}$, 188 mg/cm$^{2}$, respectively /
$-\cdot - \cdot - $ \hspace{0.2cm} the line corresponding to the
tensor polarization $p_{zz} = 0.1$, which does not depend on the
deuteron energy ).} \label{line58}
\end{figure}

\begin{figure}[!h]
\includegraphics[width=9cm,keepaspectratio]{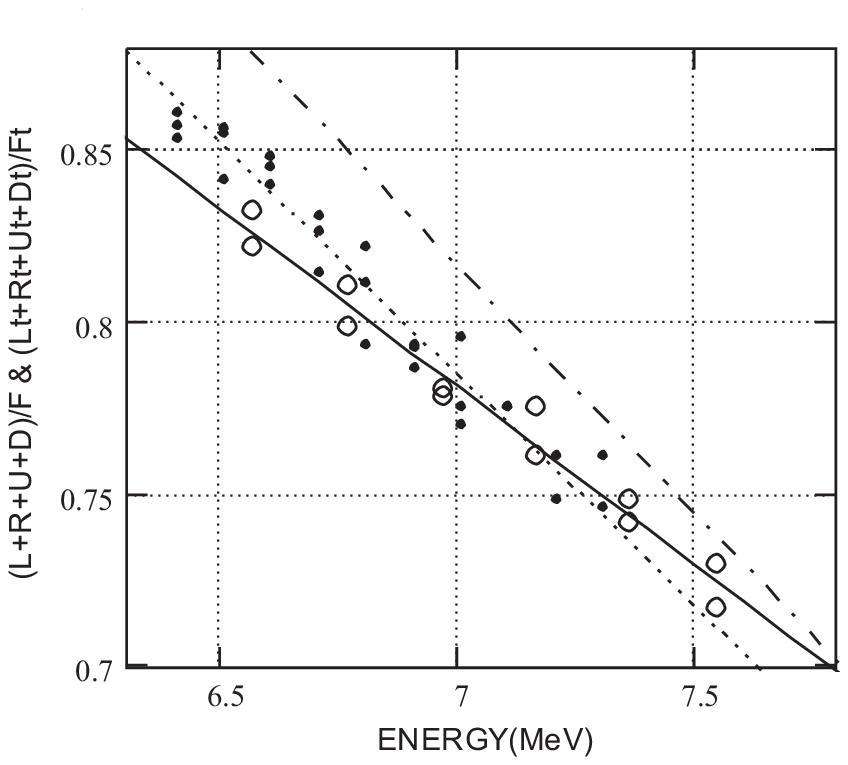}
\caption{Energy dependence of (L+R+U+D)/F without target and
(L$_{t}$+R$_{t}$+U$_{t}$+D$_{t}$)/F$_{t}$ for the target of 151
mg/cm$^{2}$ and the corresponding linear approximations (see
Fig.\@ 11).} \label{line151}
\end{figure}

\begin{figure}[!h]
\includegraphics[width=9cm,keepaspectratio]{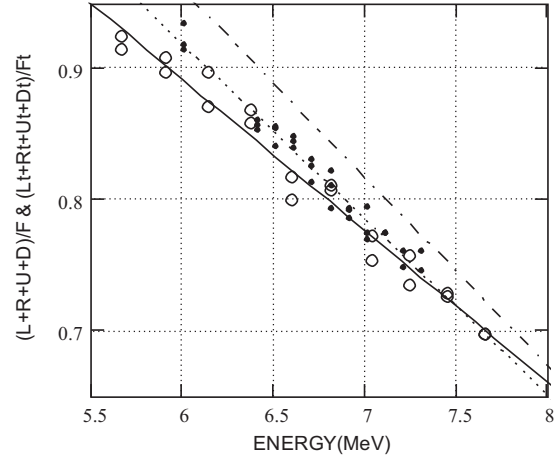}
\caption{Energy dependence of (L+R+U+D)/F without target and
(L$_{t}$+R$_{t}$+U$_{t}$+D$_{t}$)/F$_{t}$ for the target of 188
mg/cm$^{2}$ and the corresponding linear approximations (see
Fig.\@ 11).} \label{line188}
\end{figure}

Experimental points for each target and related linear
approximations are presented in Figs. \ref{line58}-\ref{line188}.
The sign $\bullet $ mark experimental points (L+R+U+D)/F without a
target (calibration) / $\circ $ \hspace{0.2cm} experimental points
(L+R+U+D)/F for the targets of 58 mg/cm$^{2}$, 151 mg/cm$^{2}$,
188 mg/cm$^{2}$, respectively /
 $\cdot \cdot \cdot \cdot \cdot $ \hspace{0.2cm} the linear fit to experimental points (L+R+U+D)/F
without target /----------- \hspace{0.2cm} the linear fit to
experimental points (L+R+U+D)/F for the targets of 58 mg/cm$^{2}$,
151 mg/cm$^{2}$, 188 mg/cm$^{2}$, respectively, the line/ $-\cdot
- \cdot - $ \hspace{0.2cm}  corresponds to the tensor polarization
$p_{zz} = 0.1$, which does not depend on the deuteron energy.


From the figures and the last table it is seen that straight lines
for all cases considered do not coincide. Especially, it is
evident for the targets of 151 mg/cm$^{2}$ and 188 mg/cm$^{2}$,
i.e.\@ we have different tensor polarization (dichroism) for
various target thickness. From the figures one more important
conclusion follows: the straight line corresponding to tensor
polarization 0.1 crosses the straight line corresponding to the
unpolarized beam of deuterons with no target in the region of 16
MeV. Experimental data behave very differently, i.e. we have a
rather strong dependence of the tensor polarization on the
deuteron energy:

\begin{itemize}

\item Experimental points have a relatively wide spread with and without
 a target, although with a target the influence of different focusing conditions
should be smaller.

 \item Since the target thickness has an error of $\pm 2\%$, and there is
 no possibility to calibrate detectors, the error of the deuteron energy after
 the target obtained with the Bethe-Bloch formula and tables is
$\pm 0.2\ MeV$. This means that the linear relation may
 be shifted in this interval for each target.
\end{itemize}

\begin{figure}[!h]
\includegraphics[width=9cm,keepaspectratio]{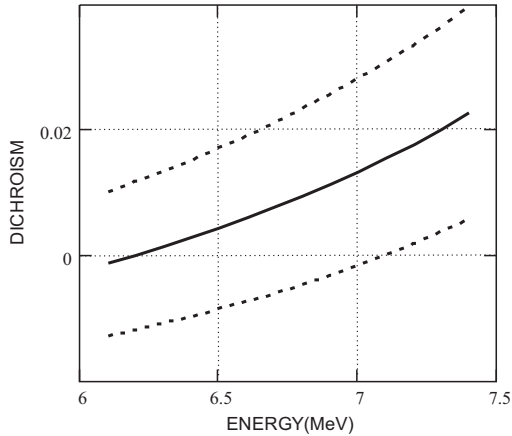}
\caption{ Dependence of dichroism on the deuteron energy after the
passage of a carbon target of 58 mg/cm$^{2}$ ( - - - -
\hspace{0.2cm} range of deviation).} \label{dich58}
\end{figure}

\begin{figure}[!h]
\includegraphics[width=9cm,keepaspectratio]{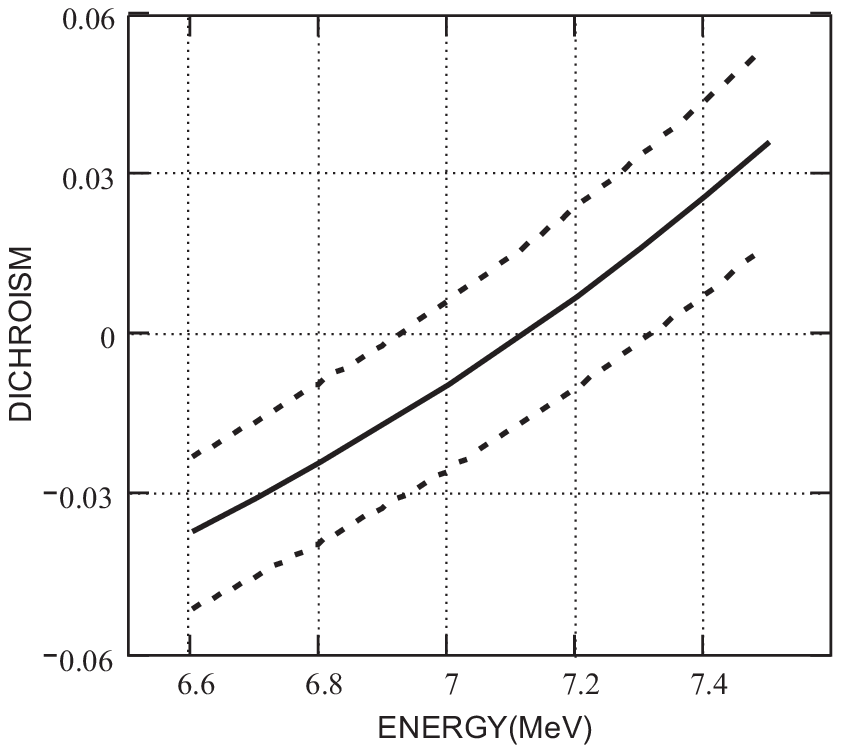}
\caption{ Dependence of dichroism on the deuteron energy after the
passage of a carbon target of 151 mg/cm$^{2}$ (see Fig.\@ 14).}
\label{dich151}
\end{figure}

\begin{figure}[!h]
\includegraphics[width=9cm,keepaspectratio]{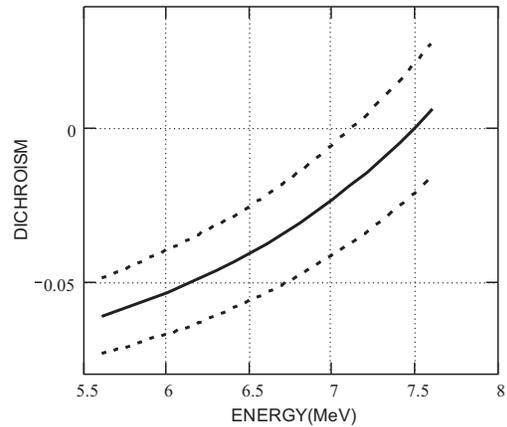}
\caption{ Dependence of dichroism on the deuteron energy after the
passage of a carbon target of 188 mg/cm$^{2}$ (see Fig.\@ 14).}
\label{dich188}
\end{figure}

Therefore, a unique source of
 information about dichroism (tensor polarization arising) is the slope
 of the straight line which does not change essentially in this error interval,
i.e. the change of line slope with and without a target
   testifies to dichroism (tensor polarization)
 arising according to (\ref{apline}). But since we do not know the
 deuteron energy after the targets exactly, we cannot find the difference between
 points on the lines corresponding to the same energy and
 analyzing power with sufficient accuracy.
 Though the magnitude of dichroism
cannot be determined
 precisely by Eq. (\ref{apline})  within its errors yet from the slope we can conclude that
dichroism exists and increases with energy in this energy region.
The average dependence of dichroism on
 the deuteron energy after the target passage is shown in
Figs.\ref{dich58}-\ref{dich188} where the relation between tensor
polarization and dichroism is used:
 $A=-\frac{{3}}{{4}}p_{zz}$.


\section{CONCLUSION}

According to experimental results it is possible to draw the
following conclusions:
\begin{enumerate}
\item For all three carbon targets spin dichroism of deuterons is
observed, especially visible for the targets of 151 mg/cm$^{2}$,
188 mg/cm$^{2}$.

\item Dichroism grows in the energy region 6-20 MeV.

\item The sign change of spin dichroism and its not being
proportional to the target thickness can be explained by a more
complicated dependence of the cross section on the energy of
particles, which changes when the beam passes through the target.
For a more complete description of the dichroism observed in this
energy region it is necessary to take into account spin-spin and
Coulomb interactions and a possible influence of resonant
reactions of deuterons with carbon.

\end{enumerate}

%References

\end{document}